\documentclass[12pt]{article}
\usepackage{epsf,latexsym}
\usepackage[all]{xy}
\usepackage{amsfonts,amssymb} %%amsmath
\epsfverbosetrue
\newcommand{\de}{\hbox{\rm{d}}}

\newcommand{\bb}{\begin{eqnarray}}
\newcommand{\ee}{\end{eqnarray}}
\newcommand{\eee}{\nonumber\end{eqnarray}}
\newcommand{\qq}{\quad}

\begin{document}

\font\twelve=cmbx10 at 13pt
\font\eightrm=cmr8

\thispagestyle{empty}

\begin{center}
${}$
\vspace{3cm}

{\Large\textbf{Cosmological constant and time delay}} \\

\vspace{2cm}

{{\large Thomas Sch\"ucker\footnote{also at Universit\'e de Provence, Marseille,
France, thomas.schucker@gmail.com } (CPT\footnote{Centre de Physique
Th\'eorique\\\indent${}$\qq\qq CNRS--Luminy, Case
907\\\indent${}$\qq\qq 13288 Marseille Cedex 9,
France\\\indent${}$\qq
Unit\'e Mixte de Recherche (UMR 6207) du CNRS et des Universit\'es
Aix--Marseille 1 et 2\\
\indent${}$\qq et Sud Toulon--Var, Laboratoire affili\'e \`a la
FRUMAM (FR 2291)}), Noureddine Zaimen\footnote{also at
Universit\'e  M'Hamed Bouguerra, 35000 Boumerdes,  Algeria, zaimennoureddine@yahoo.fr }
(LPT\footnote{
Laboratoire de Physique Th\'eorique\\\indent${}$\qq\qq
Universit\'e Mentouri\\\indent${}$\qq\qq
25000 Constantine, Algeria
}) }}

\vspace{3cm}

{\large\textbf{Abstract}}
\end{center}
The effect of the cosmological constant on the time delay caused by an isolated spherical mass is calculated without using the lens equation and compared to a recent observational bound on the time delay of the lensed quasar SDSS J1004+4112.

\vspace{2cm}

\noindent PACS: 98.80.Es, 98.80.Jk\\
Key-Words: cosmological parameters -- lensing
\vskip 1truecm

\noindent CPT-P001-2008\\
\noindent 0801.xxxx
\vspace{1cm}

${}$

\section{Introduction}

Time delay is one of the four classical tests of general relativity. The first experimental confirmation is due to I. I. Shapiro. With his team, he measured in 1968 a time delay of 240 $\mu $s for a (one-way) travel-time of 10 minutes between Mercury and Earth. In october  2007 Fohlmeister et al. \cite{fo} published a lower bound of  5.7 years on a time delay of truly cosmological nature: the travel-time between the quasar SDSS J1004+4112 and Earth is roughly $10^{10}$ years, the lens is a cluster of about $5\cdot 10^{13}$ solar masses. One month earlier Rindler \& Ishak \cite{ri} corrected the wide held error that the cosmological constant does not change the deflection angles of light. Sereno \cite{se} backed up this claim and also derived a formula for the time delay with cosmological constant. Earlier results on this time delay are due to  \cite{ke,ju}. Note also the analysis by Bakola et al. \cite{ba} on extreme lensing by black holes including a positive cosmological constant. The aim of this paper is to combine the two news and to compute the time delay between the best aligned images of the quasar (C and D) as a function of the cosmological constant. We assume the idealisation that the cluster is static and spherically symmetric and that it dominates all other lensing masses in the universe. The fact that there are five images of the quasar of course tells us that the cluster is not spherical. 

\section{The set up}

Consider a universe, which is empty except for one static, spherical, non-rotating mass $M$, the lens $L$. A source $S$, at rest with respect to the lens, emits photons, which are observed at nostra terra $T$ also assumed at rest. We neglect the masses of the source and of our local group. We use polar coordinates $(r,\theta ,\varphi )$ centered at the lens.  Because of spherical symmetry the photons' trajectory is in a plane that we take to be $\theta =\pi /2$. The angle $\varphi $ is measured with respect to the axis defined by the source, $\varphi _S=0$. With a cosmological constant, the gravitational field outside the mass $M$ is given by the Kottler metric,
\bb \de \tau ^2=B\,\de t^2-B^{-1} \de r^2-r^2\de\varphi ^2,\qq \theta =\pi /2,\qq 
B=1-\,\frac{2GM}{r} -{\textstyle\frac{1}{3}} \Lambda r^2.\ee
The source has polar coordinates $(r_S, 0)$, the earth is at $(r_T,\varphi _T)$.  Let $\alpha $ and $\alpha '$ be the two physically measured angles between the images and the cluster center and denote by $r_0$ and $r_0'$ the peri-lenses of the two light rays. We write $t_S$ and $t_T$ for the coordinate times of flight from the source to the peri-lens and from the peri-lens to Earth, see figure 1.

\begin{center}
\begin{tabular}{c}

\xy
(0,0)*{}="L";
(60,0)*{}="S";
(-45,13)*{}="T";
(-45,13)*{\bullet};
(60,0)*{\bullet};
(62.5,-3)*{S};
(-47,9.5)*{T};
(-4,-2.5)*{L};
(0,0)*{\bullet};
{\ar (0,0)*{}; (70,0)*{}}; 
"S"; "T" **\crv{(0,20)};
"S"; "T" **\crv{(-10,-20)};
"L"; "T" **\dir{-}; 
"L"; (5,13)*{} **\dir{-}; 
"L"; (-2,-6.7)*{} **\dir{-}; 
(-33,9.6)*{}; (-31.5,14.5)*{} **\crv{(-30.5,12)};
(-29,8.4)*{}; (-31.5,3.3)*{} **\crv{(-29,5)};
(-25,11)*{\alpha  };
(-24,3)*{\alpha '};
(5,0)*{}; (-4.9,1.4)*{} **\crv{(1,7)};
(6,6)*{r_0 };
(2,-4)*{r'_0};
(73,0)*{x};
(0,-14)*{};
(25,2.2)*{r_S};
(-15,6.7)*{r_T};
(-2.5,5.5)*{\varphi _T};
(32,11)*{t_S};
(32,-9)*{t'_S};
(-17,18)*{t_T};
(-21,-5.5)*{t'_T};
\endxy

\end{tabular}\linebreak\nopagebreak
{Figure 1: A double image}
\end{center}

We will suppose that $\Lambda r^2/3< 9/10$ to avoid the coordinate singularity at the equator of the de Sitter sphere.
 We will also suppose that $\delta := GM/r_0\ll 1$, $\alpha \ll 1$ and likewise for their primed quantities
 and keep only terms linear in these four quantities. Let us anticipate that in this approximation, $\alpha \sim
\sqrt{1-\Lambda r_T^2/3}\,r_0/r_T$. We will also assume that $\lambda :=\sqrt{\Lambda /3} r_0\ll 1$ and only keep terms linear in this and its primed quantity. For the example of the quasar SDSS J1004+4112, all six quantities are of the order of $10^{-5}$.

Our aim is to compute the proper time delay $\Delta \tau=\sqrt{B(r_T)}\,(t'_T+t'_S-t_T-t_S)$ as a function of $M,\ \alpha ,\ \alpha ',\ r_T,\ r_S,$ and $\Lambda $.

\section{Integrating the geodesics}

We start with the list of the non-vanishing Christoffel symbols for the Kottler metric with $\theta =\pi /2$ and denote $':=\de/\de r$,
\bb {\Gamma ^t}_{tr}=B'/(2B), &
{\Gamma ^r}_{tt}=BB'/2, &
{\Gamma ^r}_{rr}=-B'/(2B),\\
{\Gamma ^r}_{\varphi \varphi }=-rB, &
{\Gamma ^\varphi }_{r\varphi }=1/r.\ee
The geodesic equations read:
\bb &&\ddot t+B'/B\,\dot t\dot r=0,\\
&&
\ddot r+{\textstyle\frac{1}{2}} BB'\dot t^2-{\textstyle\frac{1}{2}} BB'\dot r^2-rB\dot \varphi ^2=0,\\
&&
\ddot \varphi +2r^{-1}\dot r\dot
\varphi =0,
\ee
where we denote the affine parameter by $p$ and $\dot {}:=\de/\de p$. We immediately get three first integrals:
\bb \dot t&=&1/B(r),\\
\dot\varphi &=&\frac{r_0}{r^2\sqrt{B(r_0)}},\\ 
\dot r&=&\left( 1-\,\frac{r_0^2}{r^2} \,\frac{B(r)}{B(r_0)}\right) ^{1/2} .\ee
Eliminating the affine parameter we get:
\bb\frac{\de \varphi }{\de r}&=& \pm\,
\frac{1}{r\sqrt{r^2/r_0^2-1}}\,\left[ 1-
\,\frac{2GM}{r}\,-\,\frac{2GM}{r_0}\,\frac{r}{r+r_0}\right]^{-1/2},\label{geom}\\
\frac{\de t }{\de r}&=& \pm\,
\frac{\sqrt{B(r_0)}}{B(r)\sqrt{1-r_0^2/r^2}}\,\left[ 1-
\,\frac{2GM}{r}\,-\,\frac{2GM}{r_0}\,\frac{r}{r+r_0}\right]^{-1/2}.\label{time}
\ee
Integrating equation (\ref{geom}) we obtain $\alpha \sim
\sqrt{1-\Lambda r_T^2/3}\,r_0/r_T$ and \cite{ts}
\bb \frac{r_T}{r_S} \,\sim\,\frac{4GM}{\alpha \alpha 'r_T}\,(1-\Lambda r_T^2/3)-1\label{meet}.\ee
 Let us integrate equation (\ref{time}):
 \bb t_T&=&\sqrt{B(r_0)}\int_{r_0}^{r_T}
 \frac{1}{B(r)\sqrt{1-r_0^2/r^2}}\,\left[ 1-
\,\frac{2GM}{r}\,-\,\frac{2GM}{r_0}\,\frac{r}{r+r_0}\right]^{-1/2}\de r \cr\cr
&\sim&r_0\,\sqrt{B(r_0)}\,\left[ I_{T1}+\delta\, I_{T2}
+\delta\, I_{T3}+2\,\delta\, I_{T4}\right] .\ee
We have set  $y:=r_0/r,$ $ \epsilon _T:=r_0/r_T\, >\, \lambda $,
\bb I_{T1}&:=&
\int_{\epsilon_T}^1 \frac{1}{y^2-\lambda ^2}\,\frac{\de y}{\sqrt{1-y^2}}=
\,\frac{1}{\lambda \sqrt{1-\lambda ^2}}\,
{\rm
arctanh}\left( \frac{\lambda }{\sqrt{1-\lambda ^2}}\,\frac{\sqrt{1-\epsilon_T^2}}{\epsilon_T}\right) , 
\\
I_{T2}&:=&
\int_{\epsilon_T}^1 \frac{y}{y^2-\lambda ^2}\,\frac{\de y}{\sqrt{1-y^2}}=
\,\frac{1}{\sqrt{1-\lambda ^2}}\,
{\rm
arctanh} \sqrt{\frac{{1-\epsilon_T^2} }{{1-\lambda ^2}}},\\
I_{T3}&:=&
\int_{\epsilon_T}^1 \frac{1}{y^2-\lambda ^2}\,\frac{1}{1+y} \,\frac{\de y}{\sqrt{1-y^2}}\cr&=&
\frac{1}{2(1-\lambda ^2)}\,\left[ 
2\,\frac{\sqrt{1-\epsilon_T^2}}{1+\epsilon_T} \right.\cr
&&-\,\frac{1}{\sqrt{1-\lambda ^2}} \,\ln
\frac{\left( 1+\sqrt{1-\lambda ^2}\sqrt{1-\epsilon_T^2}-\lambda \epsilon_T\right)
\left( 1+\sqrt{1-\lambda ^2}\sqrt{1-\epsilon_T^2}+\lambda \epsilon_T\right) }{\epsilon_T^2-\lambda ^2}  \cr
&&\left. -\,\frac{1}{\lambda \sqrt{1-\lambda ^2}} \,\ln
\frac{\left( 1+\sqrt{1-\lambda ^2}\sqrt{1-\epsilon_T^2}+\lambda \epsilon_T\right)\left( \epsilon_T-\lambda \right) }{  \left( 1+\sqrt{1-\lambda ^2}\sqrt{1-\epsilon_T^2}-\lambda \epsilon_T\right)\left( \epsilon_T+\lambda \right)
}\right] ,\\
I_{T4}&:=&
\int_{\epsilon_T}^1 \frac{y^3}{(y^2-\lambda ^2)^2}\,\frac{\de y}{\sqrt{1-y^2}}\cr&=&
\,\frac{2-\lambda ^2}{2\sqrt{1-\lambda ^2}^3}\,
{\rm
arctanh} \sqrt{\frac{{1-\epsilon_T^2} }{{1-\lambda ^2}}}\,+\,\frac{\lambda ^2}{2(1-\lambda ^2)}\,
\frac{\sqrt{1-\epsilon_T^2}}{\epsilon_T^2-\lambda ^2} .
 \ee
 To compute $t'_T-t_T$ we have to subtract huge numbers that are almost identical. Therefore we develop the relevant differences of the four integrals separately. Setting $x:=\alpha '/\alpha \sim r_0'/r_0$ we have:
\bb \Delta _{T1}&:=&r'_0\sqrt{B(r'_0)}I'_{T1}-
r_0\sqrt{B(r_0)}I_{T1}
\\[1mm]\sim
GM &&\!\!\!\!\!\!\!\!\!\!\!\!\!\!\!\!\left[-(1/x-1)\,\frac{{\rm
arctanh}(\lambda /\epsilon_T)}{\lambda }\, +{\textstyle\frac{1}{2}} (1-x^2)\,\frac{\epsilon_T}{\delta } \,-{\textstyle\frac{1}{2}} (1/x^2-1)\delta \,\frac{{\rm
arctanh}(\lambda /\epsilon_T)}{\lambda }\right]\nonumber ,\\[1mm]
\Delta _{T2}&:=&\sqrt{B(r'_0)}I'_{T2}-
\sqrt{B(r_0)}I_{T2}
\sim
 \ln x ,\\[1mm]
\Delta _{T3}&:=&\sqrt{B(r'_0)}I'_{T3}-
\sqrt{B(r_0)}I_{T3}\nonumber 
\\[1mm]&\sim&
(1/x-1)\,\frac{{\rm
arctanh}(\lambda /\epsilon_T)}{\lambda }\, +\ln x- (1/x^2-1)\delta \,\frac{{\rm
arctanh}(\lambda /\epsilon_T)}{\lambda },\\[1mm]
\Delta _{T4}&:=&\sqrt{B(r'_0)}I'_{T4}-
\sqrt{B(r_0)}I_{T4}
\sim
 \ln x .
\ee
The huge number $(1/x-1){\rm
arctanh}(\lambda /\epsilon_T)/\lambda$ cancels  when we add the four terms:
\bb t'_T-t_T&\sim&\Delta _{T1}+GM\Delta _{T2}+GM\Delta _{T3}+2\,GM\Delta _{T4}\nonumber\\
&\sim& GM\left[ {\textstyle\frac{1}{2}} (1-x^2)\,\frac{\epsilon_T}{\delta } \,-{\textstyle\frac{3}{2}} (1/x^2-1)\delta \,\frac{{\rm
arctanh}(\lambda /\epsilon_T)}{\lambda }\,-2\ln x\right].\ee
Finally  we have the time delay to leading order in $\delta ,\ \epsilon_\cdot$ and $\lambda $,
\bb &&\Delta \tau\sim\sqrt{1-\Lambda r_T^2/3}\,GM
\left[ 
{\textstyle\frac{1}{2}\,}\frac{\alpha ^2-\alpha '^2}{1-\Lambda r_T^2/3}\,\frac{r_T}{GM} \,
\left( 1+\,\frac{r_T}{r_S} \right) \right.\nonumber \\ && \qq
-{\textstyle\frac{3}{2}}\,  
(1-\frac{\Lambda}{3}  r_T^2)\left( \frac{1}{\alpha '^2}-\frac{1}{\alpha ^2} \right)\frac{GM}{\sqrt{\Lambda /3}\,r_T^2}
\left( {\rm
arctanh} \sqrt{\frac{\Lambda}{3}}\,r_T +
{\rm
arctanh} \sqrt{\frac{\Lambda}{3}}\,r_S \right)
 \nonumber \\ &&\qq \left.+4\ln\frac{\alpha }{\alpha '} \right] .\ee 

\section{SDSS J1004+4112}

Consider the lensing cluster of SDSS J1004+4112 and the quasar as source \cite{sh,ot}. As we have at least 4 images, the cluster cannot be spherically symmetric. We will close our eyes to the images with small $\alpha $  and consider only the images C and D with $\alpha =10''\ \pm 10\ \%$ and $\alpha' =5''\ \pm 10\ \%$. We argue that they are less senitive to a non-spherical inner struture of the cluster. The mass of the cluster is $M=(1\pm 0.2)\cdot10^{44}$ kg. The cluster has a redshift of $z_L=0.68$. A numerical integration of the Cold Dark Matter model with $\Lambda = 1.5 \cdot 10^{-52}\ {\rm m}^{-2}$ yields an area distance $d_L=7.0\cdot 10^{25}$ m as seen from Earth from. This area distance coincides with the coordinate distance $r_T$. For the quasar we have $z_S=1.734$ yielding the area distance $d_S=13.7\cdot 10^{25}$ m. Due to the magnification of the quasar by the cluster, the translation from its area distance to its coordinate distance is ambiguous. We will use two Ans\"atze,
\bb d_S=\frac{r_T+r_S}{\sqrt{1-\Lambda r_S^2/3}}\qq{\rm or}\qq
\label{var}
d_S={r_T+r_S} .
\ee 
The first one gives $r_S=5.3\cdot10^{25}$ m,
$\Lambda = 2.2 \cdot 10^{-52}\ {\rm m}^{-2}$ from equation (\ref{geom}) and $\Delta \tau=20.1$ years.  The second gives $r_S=6.7\cdot10^{25}$ m, $\Lambda = 2.7 \cdot 10^{-52}\ {\rm m}^{-2}$ and $\Delta \tau=17.1$ years. The errors on these quantities coming from the errors in $M,\ \alpha $ and $\alpha '$ can be read from tables 1 and 2,
\bb 
\Lambda =& (2.0\pm 1.4) \cdot 10^{-52}\ {\rm m}^{-2}&{\rm or}\qq(2.4\pm 1.5) \cdot 10^{-52}\ {\rm m}^{-2},\label{lam}\\
\Delta \tau=&(20.2\pm 7.5)\ {\rm years}\qq\qq
&{\rm or}\qq(18.1\pm 6.1)\ {\rm years}.\label{tau}\ee
\begin{table}
\begin{center}  
\begin{tabular}{|c||c|c|c|c|c|c|c|c|}
\hline
$M\pm 20\%$&$ +$&$-$&$-$&$-$&$-$&+&+&$+$ \\ 
\hline
$\alpha \pm 10\%$& +&+&$-$&$-$&+&+&$-$&$-$
                   \\ 
                   \hline
$\alpha' \pm 10\%$& +&+&+&$-$&$-$&$-$&$-$&+
                   \\ 
\hline\hline
$r_S\, [10^{25}\,{\rm m}]$ & 5.3&{\bf 6.1}&5.6&5.3&5.6&5.0&{\bf 4.8}&5.0  \\ 
\hline
$\Lambda [10^{-52}\,{\rm m}^{-2}]$ & 2.2&{\bf 0.7}&1.5&2.2&1.5&2.8&{\bf 3.4}&2.8    \\ 
\hline
$\Delta \tau\, [{\rm years}]$ & 24.1& 20.3&{\bf 12.8}&16.1&23.6&{\bf 27.7}& 18.7&15.1  \\
\hline
\end{tabular} 
\end{center}
\caption{Values of $r_S$ from the first of equations (\ref{var}), $\Lambda $ and $\Delta \tau$ at the corners of the error box in $M,\ \alpha $ and $\alpha '$. Minimal and maximal values are bold face.}
\end{table} 

\begin{table}
\begin{center}  
\begin{tabular}{|c||c|c|c|c|c|c|c|c|}
\hline
$M\pm 20\%$&$ +$&$-$&$-$&$-$&$-$&+&+&$+$ \\ 
\hline
$\alpha \pm 10\%$& +&+&$-$&$-$&+&+&$-$&$-$
                   \\ 
                   \hline
$\alpha' \pm 10\%$& +&+&+&$-$&$-$&$-$&$-$&+
                   \\ 
\hline\hline
$\Lambda [10^{-52}\,{\rm m}^{-2}]$ & 2.7&{\bf 0.9}&1.9&2.7&1.9&2.3&{\bf 3.9}&3.3    \\ 
\hline
$\Delta \tau\, [{\rm years}]$ & 21.8& 19.6&{\bf 12.0}&14.6&22.0&{\bf 24.2}& 16.0&13.3  \\
\hline
\end{tabular} 
\end{center}
\caption{Values of  $\Lambda $ and $\Delta \tau$ for $r_S=d_S-r_T=6.7\cdot10^{25}$ m at the corners of the error box in $M,\ \alpha $ and $\alpha '$. Minimal and maximal values are bold face.}
\end{table} 

It is encouraging to note that both constraints (\ref{lam}) on the cosmological constant are compatible with the present observational bounds, $\Lambda = (1.5\pm 0.7) \cdot 10^{-52}\ {\rm m}^{-2}$, whose   central value we have used without variation to translate redshifts into area distances. Also both constraints (\ref{tau}) on the time delay are compatible with the recent observational lower bound \cite{fo} $\Delta \tau >5.7$ years.

\section{Conclusions}

Our computation rests on three shaky idealisations:
\begin{itemize} \item
the sphericity of the cluster, 
\item
ignoring the velocity of observer and source with respect to the lens,
\item
ignoring all masses except for the cluster.
\end{itemize}
The first raises the good old question whether a spherical cow can be useful. To go beyond it implies to take into account multiple scattering of the photon off sub-constituents of the cluster. Calculating angles,
 time delays and number of images from the density profile of the cluster and its dark matter halo is a well developed art \cite{eh}. Applying it to SDSS J1004+4112, Kawano \& Oguri \cite{ko}            
predict a time delay between images C and D of up to 10 years. However including a positive cosmological constant in their analysis is not straight forward.

Khriplovich \& Pomeransky \cite{kh} point out that if the earth is taken comoving with respect to the exponentially expanding de Sitter space then the effect of the cosmological constant on the deflection cancels.

Going beyond the third idealisation necessitates an interpolating solution, which embeds many static, curved Kottler solutions into the ambient expanding, flat Friedmann solution. This is a long standing problem, to which already Einstein \& Straus \cite{es,sch} have contributed an unstable \cite{kr} solution.
A first qualitative assessment based on this solution is given in section III of reference \cite{ir}: the other clusters in the universe weaken the effect of the cosmological constant and increase upper bounds on the cosmological constant from certain lenses by two orders of magnitude. 

It is safe to conclude: we all agree that lensing and time delay at cosmological scale is beautiful 
physics and that our theoretical understanding of it is incomplete. For the older physicists among us however, a time delay of 10 or 20 years is bad news.
\vskip 1cm \noindent
Note added: Mustapha Ishak \cite{mu} has extended the analysis of how the embedding of Kottler's solution into Friedmann's solution modifies light deflection to include the time delay. We thank him for having kindly sent us his work.


\begin{thebibliography}{10}

\bibitem{fo}
  J.~Fohlmeister, C.~S.~Kochanek, E.~E.~Falco, C.~W.~Morgan and J.~Wambsganss,
  ``The Rewards of Patience: An 822 Day Time Delay in the Gravitational Lens
  SDSS J1004+4112,''
  arXiv:0710.1634 [astro-ph].
\bibitem{ri}
  W.~Rindler and M.~Ishak,
  ``The Contribution of the Cosmological Constant to the Relativistic Bending
  of Light Revisited,''
  Phys.\ Rev.\  D {\bf 76} (2007) 043006
  [arXiv:0709.2948 [astro-ph]].
  \bibitem{se}
  M.~Sereno,
  ``On the influence of the cosmological constant on gravitational lensing in
  small systems,''
  arXiv:0711.1802 [astro-ph].
  \bibitem{ke}
  A.~W.~Kerr, J.~C.~Hauck and B.~Mashhoon,
  ``Standard Clocks, Orbital Precession and the Cosmological Constant,''
  Class.\ Quant.\ Grav.\  {\bf 20} (2003) 2727
  [arXiv:gr-qc/0301057].
  \bibitem{ju}
  V.~Kagramanova, J.~Kunz and C.~L\"ammerzahl,
  ``Solar system effects in Schwarzschild--de Sitter spacetime,''
  Phys.\ Lett.\  B {\bf 634} (2006) 465
  [arXiv:gr-qc/0602002].
  \bibitem{ba}
  P.~Bakala, P.~\v{C}erm\'ak, S.~Hled\'{\i}k, Z.~Stuchl\'{\i}k and K.~Truparov\'a,
 ``Extreme gravitational lensing in vicinity of Schwarzschild-de Sitter black
holes,''
  arXiv:0709.4274 [astro-ph].
  \bibitem{ts}
  T.~Sch\"ucker,
  ``Cosmological constant and lensing,''
  arXiv:0712.1559 [astro-ph].
\bibitem{sh}
  K.~Sharon {\it et al.},
 ``Discovery of Multiply Imaged Galaxies behind the Cluster and Lensed Quasar
  SDSS J1004+4112,''
  Astrophys.\ J.\  {\bf 629} (2005) L73
  [arXiv:astro-ph/0507360].
\bibitem{ot}
  N.~Ota {\it et al.},
  ``Chandra Observations of SDSS~J1004+4112: Constraints on the Lensing Cluster
and Anomalous X-Ray Flux Ratios of the Quadruply Imaged Quasar,''
  Astrophys.\ J.\  {\bf 647} (2006) 215
  [arXiv:astro-ph/0601700].
\bibitem{eh}
P. Schneider, J. Ehlers and E. Falco,
``Gravitational Lenses,'' Springer (1992).
\bibitem{ko}
  Y.~Kawano and M.~Oguri,
  ``Time Delays for the Giant Quadruple Lensed Quasar SDSS J1004+4112:
  Prospects for Determining the Density Profile of the Lensing Cluster,''
  Publ.\ Astron.\ Soc.\ Jap.\  {\bf 58} (2006) 271
  [arXiv:astro-ph/0601149]. 
  \bibitem{kh}
  I.~B.~Khriplovich and A.~A.~Pomeransky,
  ``Does Cosmological Term Influence Gravitational Lensing?,''
  arXiv:0801.1764 [gr-qc]. 
\bibitem{es}
A. Einstein and E. G. Straus,
``The influence of the expansion of space on the gravitation fields surrounding the individual star,''
Rev. Mod. Phys. {\bf 17} (1945) 120, {\bf 18} (1946) 148.
\bibitem{sch}
E. Sch\"ucking,
``Das Schwarzschildsche Linienelement und die Expansion des Weltalls,''
Z.\ Phys.\  {\bf 137} (1954) 595.
\bibitem{kr}
A. Krasi\'nski,
``Inhomogeneous Cosmological Models,''
Cambridge University Press (1997),  page 113.
\bibitem{ir}
  M.~Ishak, W.~Rindler, J.~Dossett, J.~Moldenhauer and C.~Allison,
  ``A New Independent Limit on the Cosmological Constant/Dark Energy from the
  Relativistic Bending of Light by Galaxies and Clusters of Galaxies,''
  arXiv:0710.4726 [astro-ph].
  \bibitem{mu}
  M.~Ishak,
  ``Light Deflection, Lensing, and Time Delays from Gravitational Potentials
  and Fermat's Principle in the Presence of a Cosmological Constant,''
  arXiv:0801.3514 [astro-ph].

\end{thebibliography}
\end{document}